\documentclass[useAMS,usenatbib]{mn2e}   
\usepackage{psfig,epsf}   
 
\newcommand{\ltsima}{$\; \buildrel < \over \sim \;$}   
\newcommand{\lesssim}{\lower.5ex\hbox{\ltsima}}   
\newcommand{\gtrsim}{\lower.7ex\hbox{$\;\stackrel{\textstyle>}{\sim}\;$}}

\def\la{\mathrel{\mathchoice {\vcenter{\offinterlineskip\halign{\hfil   
$\displaystyle##$\hfil\cr<\cr\sim\cr}}}   
{\vcenter{\offinterlineskip\halign{\hfil$\textstyle##$\hfil\cr<\cr\sim\cr}}}   
{\vcenter{\offinterlineskip\halign{\hfil$\scriptstyle##$\hfil\cr<\cr\sim\cr}}}   
{\vcenter{\offinterlineskip\halign{\hfil$\scriptscriptstyle##$\hfil\cr<\cr\sim\cr}}}}}   
\def\ga{\mathrel{\mathchoice {\vcenter{\offinterlineskip\halign{\hfil   
$\displaystyle##$\hfil\cr>\cr\sim\cr}}}   
{\vcenter{\offinterlineskip\halign{\hfil$\textstyle##$\hfil\cr>\cr\sim\cr}}}   
{\vcenter{\offinterlineskip\halign{\hfil$\scriptstyle##$\hfil\cr>\cr\sim\cr}}}   
{\vcenter{\offinterlineskip\halign{\hfil$\scriptscriptstyle##$\hfil\cr>\cr\sim\cr}}}}}   
\def\kms{{km~s$^{-1}$}}   
\def\fmag{\hbox{$.\!\!^{\rm m}$}}    
\def\degr{\hbox{$^\circ$}}   
   
\def\arcsec{\hbox{$^{\prime\prime}$}}

\def\farcs{\hbox{$.\!\!^{\prime\prime}$}}   
   
\title[NGC\,3741: dark halo profile from the most extended rotation    
curve]{NGC\,3741: dark halo profile from the most extended rotation curve}   
\author[Gianfranco Gentile, Paolo Salucci, Uli Klein and Gian Luigi Granato]{Gianfranco    
Gentile$^{1}$\thanks{E-mail: ggentile@unm.edu}, Paolo Salucci$^{2}$,    
Uli Klein$^{3}$, and Gian Luigi Granato$^{4}$ \\   
$^{1}$University of New Mexico, Department of Physics and Astronomy, 800 Yale   
Blvd NE, Albuquerque, NM 87131, USA\\   
$^{2}$SISSA, via Beirut 4, 34014 Trieste, Italy \\   
$^{3}$Argelander-Institut f\"ur Astronomie, Universit\"at Bonn, Auf dem H\"ugel 71,   
53121 Bonn, Germany\\
$^{4}$INAF - Osservatorio Astronomico di Padova, Vicolo Osservatorio 5, 35122
Padova, Italy
}   
\begin{document}   
   
\date{Accepted. Received}   
   
   
\maketitle   
   
\label{firstpage}   
   
\begin{abstract}   
   
We present new HI observations of the nearby dwarf galaxy NGC\,3741. This galaxy    
has an extremely extended HI disk, which allows us to trace the rotation curve    
out to unprecedented distances in terms of the optical disk: we reach 42 B-band    
exponential scale lengths or about 7 kpc. The HI disk is strongly warped, but the warp is very 
symmetric. The distribution and kinematics are accurately derived by building 
model data cubes, which closely reproduce the observations. In order to account 
for the observed features in the data cube, radial motions of the order of 
5$\ldots$13~km~s$^{-1}$ are needed. They are consistent with an inner bar of 
several hundreds of pc and    
accretion of material in the outer regions. 
   
The observed rotation curve was decomposed into its stellar, gaseous and dark    
components. The Burkert dark halo (with a central constant density core)   
provides very good fits. The dark halo density distribution predicted by   
the $\Lambda$CDM theory fails to fit the data, unless NGC\,3741 is a 2.5-$\sigma$    
exception to the predicted relation between concentration parameter and virial    
mass and has at the same time a high value of the virial mass 
(though poorly constrained) of 10$^{11}$ M$_{\odot}$. Noticeably, MOND seems to be     
consistent with the observed rotation curve. 
Scaling up the contribution of the    
gaseous disk also gives a good fit.  
   
\end{abstract}   
   

\section{Introduction}   
   
The measurement of rotation curves of disk galaxies has been one of the most   
powerful tools to study dark matter, its content relative to baryons, and its    
distribution. In particular, dwarf galaxies are good candidates for dark-matter    
studies as their kinematics is generally dominated by dark matter down to small    
galactocentric radii (Persic, Salucci \& Stel 1996),
 thus providing an almost clean measurement of the dark-matter    
contribution to the observed rotation curve and hence of its density profile.    
Therefore, they represent an important test for cosmological models, because    
numerical simulations performed in the framework of the currently favoured theory    
of structure formation in the Universe, $\Lambda$ Cold Dark Matter ($\Lambda$CDM),    
predict a well-defined density profile for dark matter in virialised structures,    
the NFW profile (Navarro, Frenk and White 1996). Subsequent modifications    
and refinements via improved numerical resolution (Moore et al. 1999, Navarro et al.    
2004) do not change one of the main predicted features: a central density cusp,   
$\rho(r) \propto r^{-\alpha}$ for $r \rightarrow 0$, with $1.0 < \alpha < 1.5$.    
   
In the recent past, several studies were devoted to the analysis of rotation    
curves of dwarf galaxies (e.g. van den Bosch \& Swaters 2001, Borriello \&
Salucci 2001, Swaters et al. 2003,
Weldrake, de Blok \& Walter 2003, Simon et al. 2005, Gentile et al. 2005), in particular to    
test whether the density profile predicted by $\Lambda$CDM models is actually    
observed. Even though the debate as to whether observations can put strong
constraints on the dark matter density profiles is still ongoing, the results are that in most cases the 
observations prefer    
halos with a constant-density core, $\rho(r) \sim \rho_0$ for $r \rightarrow 0$.
A possible explanation to reconcile the observations and predictions 
involves non-circular motions induced by halo triaxiality (Hayashi et 
al. 2004, Hayashi \& Navarro 2006); the authors show the effects of 
non-circular motions on the observed minor axis kinematics
for some viewing angles.
Gentile et al. (2005) analysed the velocity field of DDO 47 (of which 
the rotation curve is best fitted by a cored halo) through
its harmonic decomposition and found non-circular motions of the order
of only $2 - 3$ km s$^{-1}$, an order of magnitude smaller than the
discrepancy between the observations and the $\Lambda$CDM predictions.
However, the effects of triaxiality vary strongly with the viewing geometry.
A detailed study of the effects of triaxiality with all possible viewing 
angles on the results of harmonic decompositions of velocity fields
is yet to come and would be a valuable result. 
It seems however unlikely that the very small 2-D non-circular motions  
measured in Gentile et al.(2005) may hide an underlying NFW cusp.

NGC\,3741 is a sort of privileged laboratory to investigate the dark matter
distribution in galaxies.
It is a nearby dwarf irregular galaxy with an absolute blue magnitude of    
-13\fmag13. Its distance (Karachentsev et al. 2004), estimated through the tip of    
the Red-Giant Branch, is 3.0$\pm$0.3~Mpc,
while Georgiev et al. (1997) find a slightly higher value (3.5 $\pm$ 0.7 Mpc)
from the brightest blue stars; unless stated otherwise, in this paper we adopt a distance
of 3 Mpc, obtaining an absolute B-band blue luminosity of 2.7 $\times$ 10$^{7}$ $L_{\odot}$. 
The HI disk of this galaxy was studied    
by Begum et al. (2005) using observations performed with the Giant Metrewave   
Radio Telescope (GMRT), which disclosed the extremely extended gaseous disk of    
NGC\,3741. They traced  the rotation curve out to 38 B-band times the exponential scale    
length, whose value is 10\farcs75 (Bremnes et al. 2000),
corresponding to 0.16 kpc at a distance of 3 Mpc.  
We observed this galaxy in the HI line with the Westerbork    
Synthesis Radio Telescope (WSRT) and reach 42 exponential scale lengths. 
We re-determined the rotation curve by means    
of the more efficient Modified Envelope Tracing method (Gentile et al. 2004) 
and data cube modelling. 
This is to our    
knowledge the most extended rotation curve ever measured in terms of the    
optical size.
   
\begin{figure}   
\psfig{figure=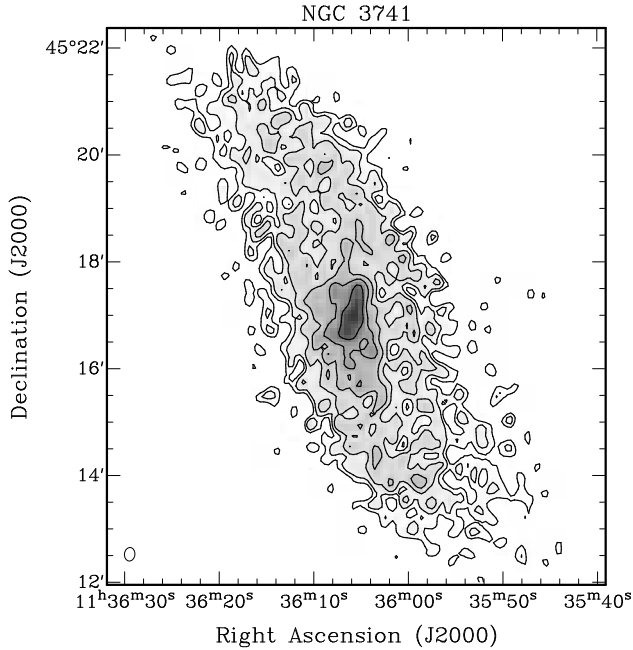,width=8.5cm,angle=-90}   
\caption{HI total intensity map from the high-resolution data cube.
Contours are (3, 4.5, 6.75, ...) $\times$ 10$^{20}$ atom~cm$^{-2}$.
The beam is shown in the bottom left corner.}
\label{mom0_only}   
\end{figure}   

\begin{figure}   
\psfig{figure=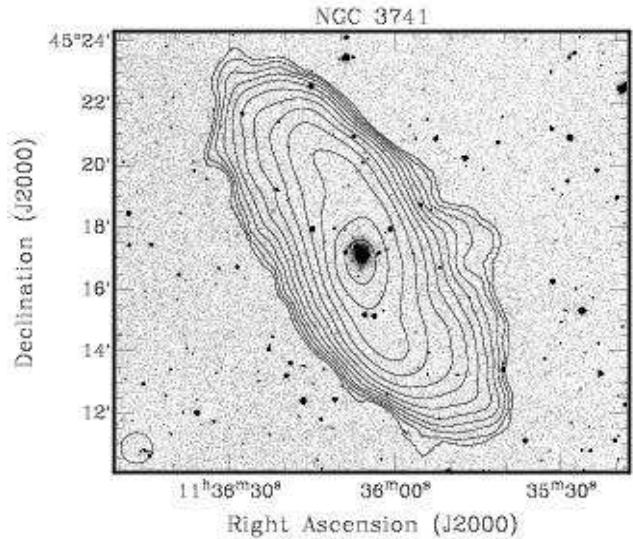,width=8.5cm,angle=-90}   
\caption{HI total intensity map from the low-resolution 
data cube (contours) superimposed
onto an optical image (DSS, greyscale).
Contours are (1.5, 2.25, 3.375, ...)  $\times$ 10$^{19}$ atom~cm$^{-2}$.
The radio beam is shown in the bottom left corner.}
\label{mom0_opt_only}   
\end{figure}

\begin{figure}   
\psfig{figure=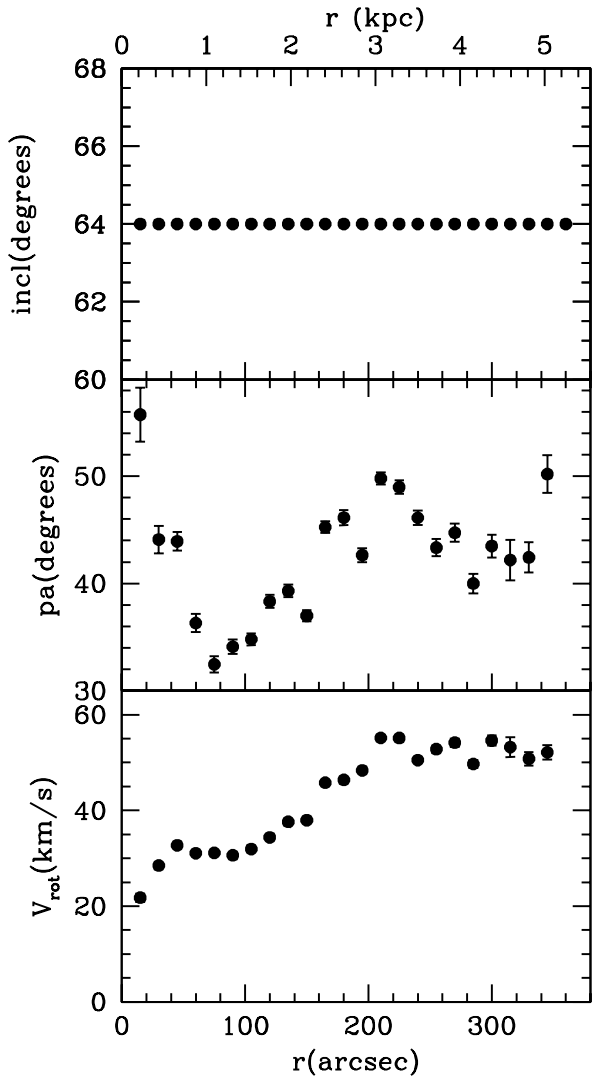,width=8cm}   
\caption{Parameters of our first attempt to derive the rotation curve of   
NGC\,3741: the tilted-ring fit to the high-resolution velocity field, with    
the inclination fixed and the position angle as a free parameter. 
}
\label{rotcur}   
\end{figure}

\begin{figure}   
\psfig{figure=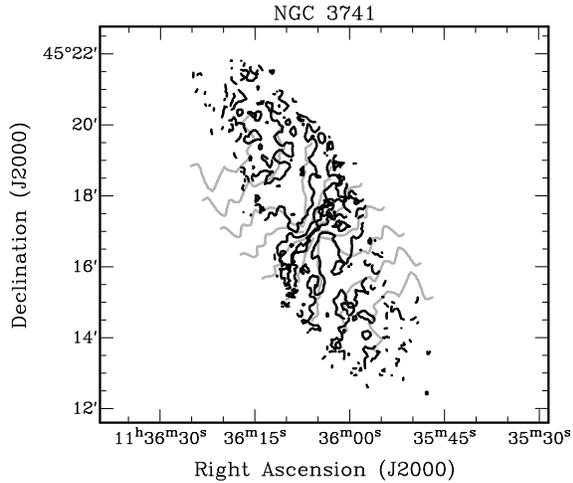,width=7.8cm,angle=0}   
\caption{Comparison between the observed velocity field derived from the    
Modified Envelope Tracing method (black contours, see text) and the modelled one (grey contours),    
built with the output position angles from the tilted-ring fit to the velocity    
field. The beam is shown in the bottom left corner. Contours are 228~km~s$^{-1}$    
(the systemic velocity), $\pm 10, 20, 30, \ldots$~\kms. }   
\label{velfi}   
\end{figure}

\begin{figure}   
\psfig{figure=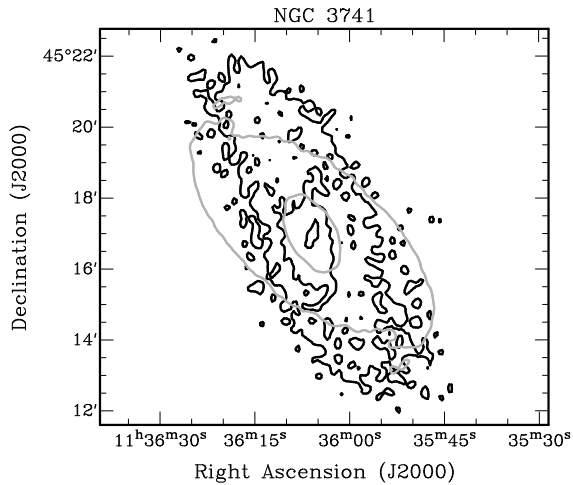,width=7.8cm,angle=0}   
\caption{Comparison between the observed high-resolution total intensity map   
and the one computed from a model data cube based on the output parameters    
from the tilted-ring fit on the velocity field. The beam is shown in the bottom    
left corner. Contours are (3, 9, 27) $\times$ 10$^{20}$ atom~cm$^{-2}$. }   
\label{mom0_velfi}   
\end{figure}   
   
\begin{figure}   
\psfig{figure=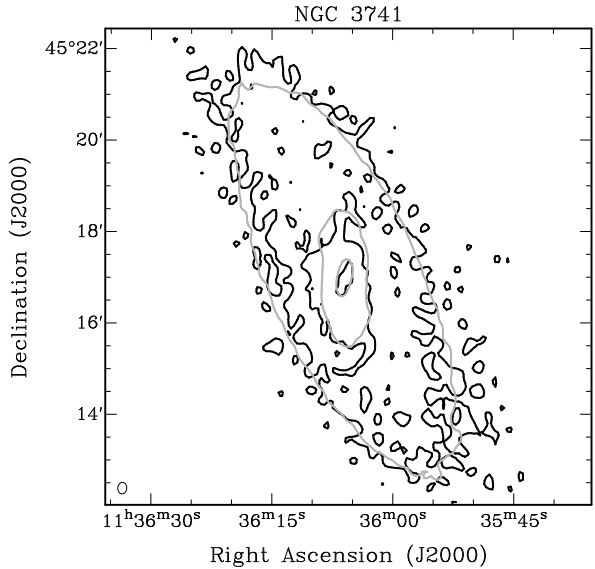,width=7.8cm,angle=-90}   
\caption{Comparison between the observed high-resolution total-intensity map and that computed    
from the model data cube, built with our final geometrical parameters. The beam    
is shown in the bottom left corner. Contours are (3, 9, 27) $\times$ 10$^{20}$    
atom~cm$^{-2}$. }   
\label{mom0}   
\end{figure}

\begin{figure}   
\psfig{figure=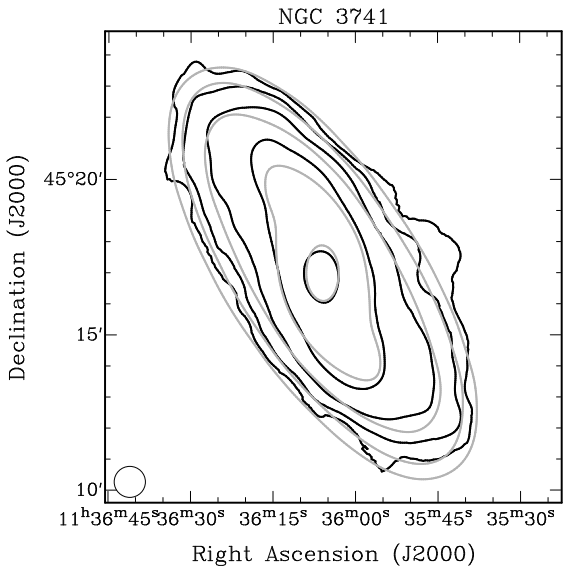,width=7.8cm,angle=-90}   
\caption{Comparison between the observed total-intensity map of the    
low-resolution cube and the one computed from the model data cube    
built with our final geometrical parameters. The beam is shown in the    
bottom left corner. Contours are (1.5, 4.5, 13.5, ...) $\times$ 10$^{19}$    
atom~cm$^{-2}$. }   
\label{mom0_lores}   
\end{figure}

\begin{figure}   
\begin{center}   
\psfig{figure=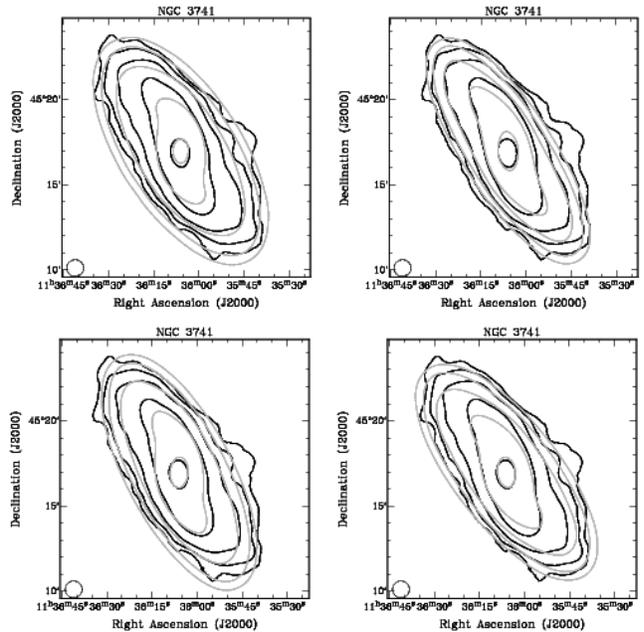,width=8.5cm,angle=0}   
\caption{   
Total intensity maps of the low-resolution cube. Observations: black contours;
Model: grey contours. Top left: using our best inclination minus 5\degr.
Top right: our best inclination plus 5\degr.
Bottom left: our best position angles minus 5\degr.
Bottom right: our best position angles plus 5 \degr.
}
\label{pa_incl}   
\end{center}   
\end{figure}

\begin{figure}   
\begin{center}   
\psfig{figure=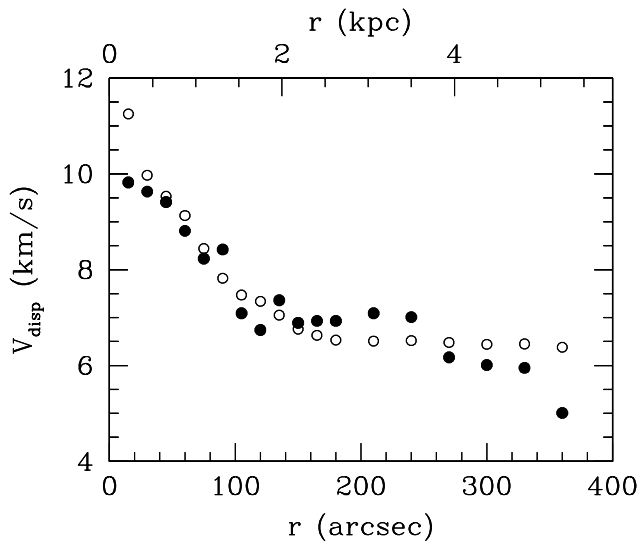,width=8.5cm,angle=0}   
\caption{   
Median values of the velocity dispersion from the velocity dispersion
maps integrated over ellipses. Full circles: from the observed high-resolution
data cube. Empty circles: from a high-resolution model data cube with a
constant velocity dispersion of 6 \kms.
}
\label{vdisp}   
\end{center}   
\end{figure}

\begin{figure}   
\psfig{figure=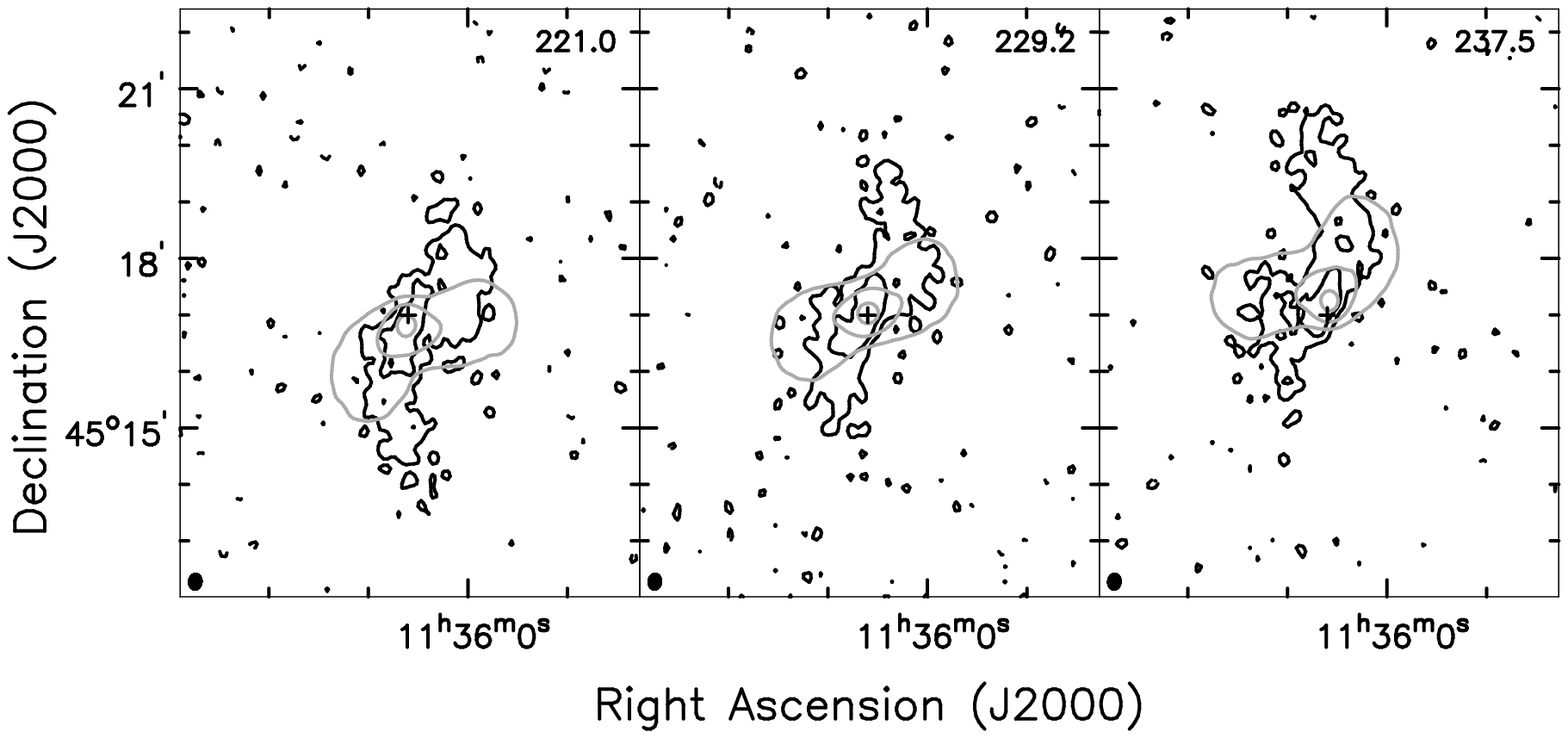,width=8.5cm,angle=-90}   
\caption{Comparison between the observed central channel maps of the high-resolution    
data cube and the modelled ones, using the same inclination and position angles as    
in Fig. \ref{mom0} and disregarding non-circular motions. Contours are -2.25, 2.25    
(2.5~$\sigma$), 6.75, and 20.25~mJy/beam. The beam is shown in the bottom left    
corner, and the cross indicates the centre of the galaxy. }   
\label{cubes_hires_vr0}   
\end{figure}

\section{Observations and reduction}   
\label{obsred}   

The observations were performed with the WSRT for 12    
hours with the `maxi-short' configuration. The correlator setup was chosen   
to have 10~MHz total bandwidth divided into 2048 channels, which were    
subsequently reduced through Hanning smoothing. The final data cubes have    
a velocity resolution of 4.1~\kms. Flagging and calibration were performed    
with the MIRIAD software. A continuum map was created from the line-free    
channels, which was used for phase self-calibration. The continuum was    
subtracted from the u-v data using line-free channels and then an HI data    
cube was created, using uniform weighting. The data were CLEANed in the    
regions of the channel maps with HI emission and subsequently restored using    
a Gaussian beam. The beam size is 15\farcs2 $\times$ 11\farcs7, which at    
the distance of NGC\,3741 corresponds to approximately 220~pc $\times$ 170~pc.    
We also created a low-resolution data cube, with a beam of 60\arcsec $\times$    
60\arcsec. 
In order to separate real and spurious emission, the data cubes were smoothed
at about twice the resolution; the regions in these smoothed cubes with emission
lower than 2$\sigma$ were blanked. 
Based on these masked data cubes, total HI maps (moment 0 maps) from the high- and
low-resolution cubes were created: they are shown in Figs. \ref{mom0_only} and 
\ref{mom0_opt_only}.

The integrated HI flux is 59.6 Jy km s$^{-1}$, which corresponds
to an HI mass of $1.3 \times 10^8$ M$_\odot$, and a total gas mass (including
primordial He) of $M_{\rm gas}=1.7 \times 10^8$ M$_\odot$. The value of 
our measured total flux is higher than the single-dish one (Schneider et
al. 1992), but it is lower than the value quoted in Begum et al. (2005).
However, it is unlikely that we have lost a significant amount of flux due 
to missing short spacings, because the largest angular scale imaged by the 
WSRT in the maxi-short configuration is much larger than the size of the 
emission in the channel maps.

\section{Data cube modelling}   
\label{modelling}   
   
\subsection{Velocity field: a first analysis}
\label{velfi_first}

From the above-mentioned data cubes at two different resolutions velocity   
fields were constructed based on the Modified Envelope Tracing method 
(Gentile et al. 2004).    
A first attempt to understand the kinematics of NGC\,3741 was made by    
performing a tilted-ring fit (Begeman 1989) to the abovementioned velocity fields 
using the task ROTCUR within GIPSY (van der Hulst et al. 1992),    
and leaving the inclination and position angles as free parameters. 
This way,   
the orientation angles were very poorly constrained. In order to obtain a    
first estimate of the radial run of the position angle, we fixed the    
inclination at the value of 64\degr, the average between the two    
extremes of the inclination range fitted by Begum et al. (2005); the    
results are shown in Fig. \ref{rotcur}. The kinematical position angles    
found this way do not agree with the morphological position angles, as    
can be seen in Figs. \ref{velfi} and \ref{mom0_velfi}. In Fig. \ref{velfi}    
we compared a model velocity field based on the parameters derived from   
the tilted-ring modelling to the observed velocity field: even though the   
observed and modelled
contours are very close to each other, the overall shape of the modelled    
galaxy does not coincide with the observed one. This points towards a    
misalignment between the morphological and kinematical position angles   
and it is a first hint at the presence of non-circular motions,
that were not taken into account by Begum et al. (2005).

\subsection{Improved orientation parameters}
\label{improved}

As we have shown in the previous section, the orientation parameters derived from the    
tilted-ring fit of the velocity field cannot account for the observed    
data cube. 
We then built synthetic data cubes (using a modified version of the task    
GALMOD in GIPSY, Barbieri et al. 2005), which were compared to the actual    
observations. These model observations are based on a number of geometrical    
and kinematical parameters that describe the HI disk, such as the rotation    
curve and the inclination and position angles as a function of radius.    
The other input parameters are: the central position and the systemic velocity,
which can be derived with reasonable accuracy from the observations;
the velocity dispersion as a function of radius, which we kept
at a fixed value of 6 km s$^{-1}$, typical of dwarf irregulars 
(Weldrake et al. 2003); the surface density as a function of radius, which
is derived from the total HI map; the scale height of the HI disk,
which we assumed to be 300 pc, a typical value for dwarf galaxies;
the radial motions as a function of radius.  
With the choice of parameters from Section \ref{velfi_first}, 
the position angle was incorrect, and the    
inclination was underestimated, as can be seen in Figs. \ref{velfi} and 
\ref{mom0_velfi}. These two parameters were then iteratively    
changed in order to give a better representation of the total HI map;    
the results are show in Figs. \ref{mom0} and \ref{mom0_lores}
for the high- and low-resolution data cubes respectively, in which the models
clearly match the data much better. Compared to the position angles shown in Fig.    
\ref{rotcur}, the new ones are different by values ranging from 70\degr\    
in the inner parts to about 15\degr\ in the outer ones. 
The inclination    
was changed to a fixed value of 70\degr, in order to obtain good agreement
with the total HI maps, both at high and low resolution. The radial dependence
of the position angle is shown in the middle panel of Fig. \ref{parametri}.
In Fig. \ref{pa_incl} we show low-resolution total intensity maps
with different choices of the orientation parameters. 
Based upon these models, we estimate the uncertainty of    
the inclination and position angles to be about 4\degr.
In the innermost few data points we conservatively doubled this value,
given the smaller number of beams that sample the total HI map
and the higher complexity of the HI kinematics and distribution.
Note that the assumption of a constant velocity dispersion
of 6 km s$^{-1}$ is consistent with the observations: even though
it is apparently increasing towards the centre, a velocity 
dispersion map of a model data cube with a constant velocity
dispersion of 6 km s$^{-1}$ exhibits a very similar behaviour
(Fig. \ref{vdisp}), because of the broadening of the velocity
profiles due to the beam where the velocity gradients are the
largest.

\subsection{Harmonic decomposition of the velocity field}
\label{section_harmonic}   

\begin{figure*}   
\begin{center}   
\hspace{1.1cm}   
\psfig{figure=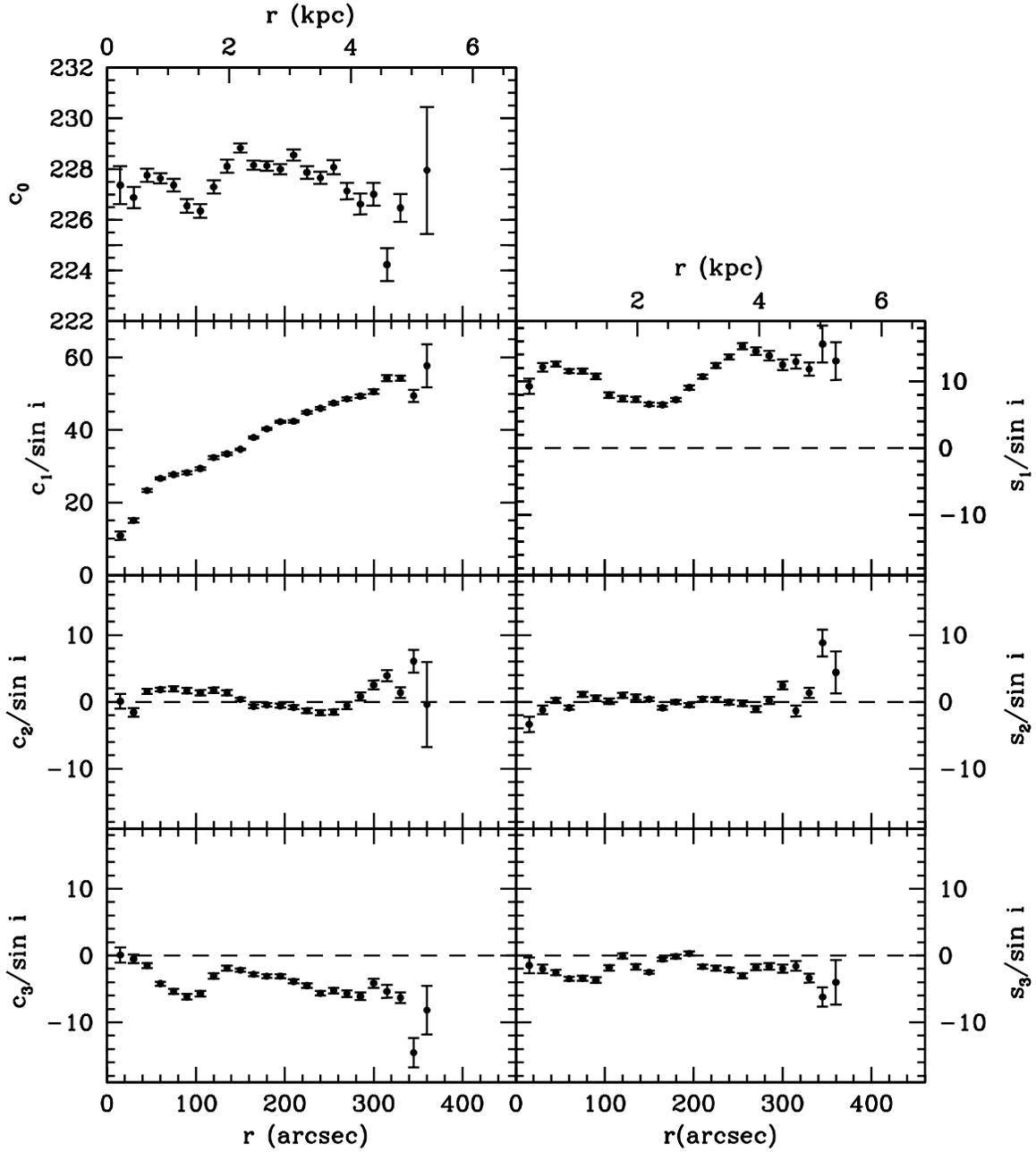,width=16cm,angle=0}   
\caption{Results of the harmonic decomposition of the high-resolution    
velocity field using the geometrical parameters based on data cube modelling.   
}
\label{harmonic}   
\end{center}   
\end{figure*}

\begin{figure}   
\begin{center}   
\psfig{figure=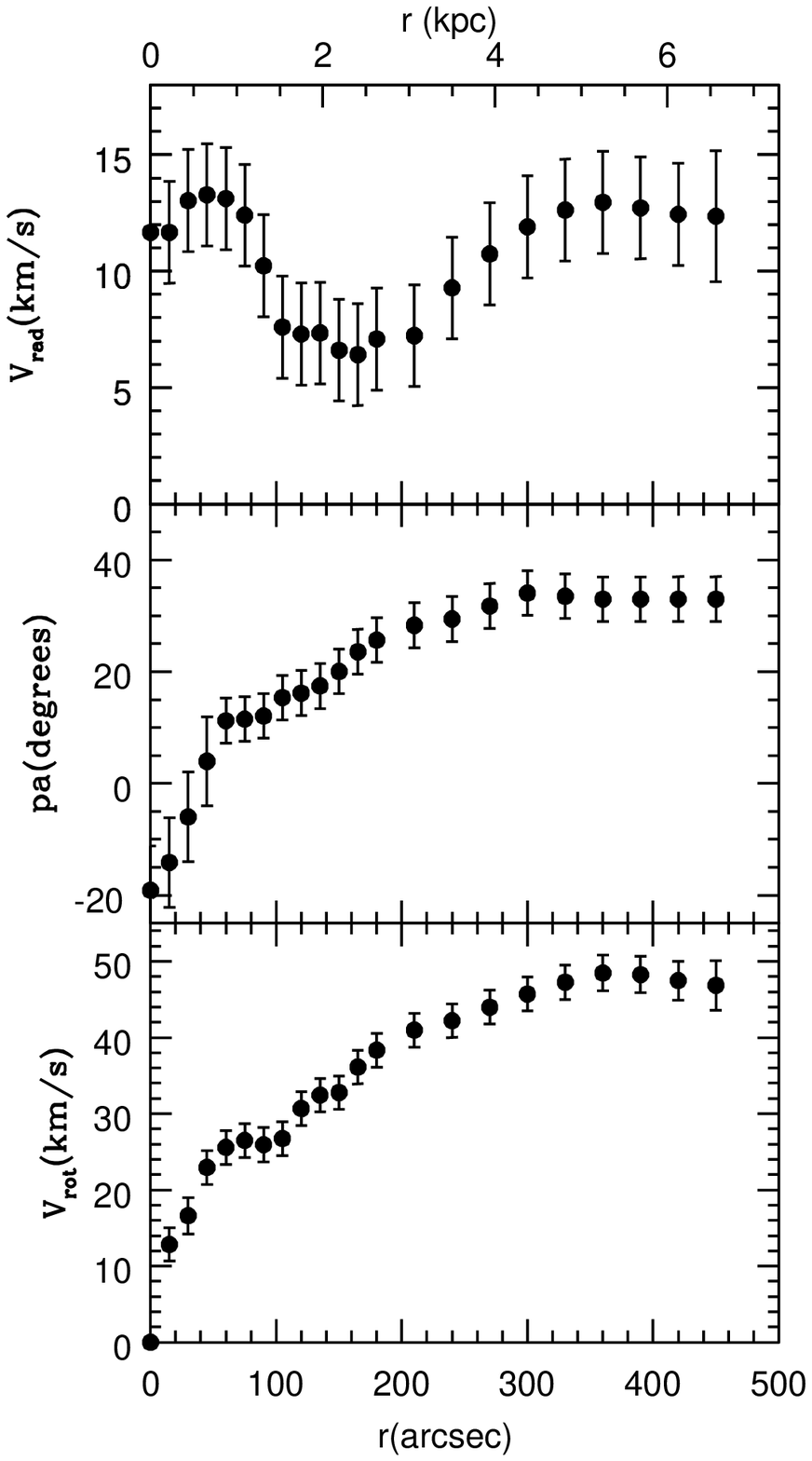,width=8.5cm,angle=0}   
\caption{   
The geometrical and physical parameters that lead to the final data cubes. 
The radial velocities are discussed in  Sections \ref{section_harmonic}, 
\ref{section_high}, \ref{model_lores} and \ref{section_non}.
The position angle is derived in Section \ref{improved} while
the rotation curve is derived in Sections \ref{section_harmonic}, 
\ref{section_high} and \ref{model_lores}. 
}   
\label{parametri}   
\end{center}   
\end{figure}

In order to account for the kinematics, and therefore to explain the    
difference between morphological and kinematical position angles, one must    
resort to non-circular motions. A data cube built with the inclination    
and position angles that match the total HI map is inconsistent with the    
observations if non-circular motions are kept to zero    
(Fig.~\ref{cubes_hires_vr0}). An estimate of their amplitude can be found    
by performing a harmonic decomposition of the velocity field (Schoenmakers    
et al. 1997, Wong, Blitz \& Bosma 2004), keeping the position and inclination angles    
fixed at the values derived in the previous Section, since leaving them as free 
parameters leads to unconstrained fits. 
With this method the observed line-of-sight velocity $V_{\rm los}$    
is modelled by    
   
\begin{equation}   
V_{\rm los} = c_0 + \sum_{j=1}^n [c_j {\rm cos}(j \psi) + s_j {\rm sin}(j \psi)],    
\end{equation}   
   
\noindent    
where $\psi$ is the azimuthal angle. In this analysis we considered    
terms up to $j=3$. The results are illustrated in Figs. \ref{harmonic} for
the high-resolution cube (the low-resolution cube gives similar results), where    
the $s_1$ component (indicative of radial motions) is non-zero and    
varies approximately between 5 and 10~km s$^{-1}$, with a region between    
150\arcsec\ and 250\arcsec\ where they are close to 5~\kms. The very small    
values of $c_2$ and $s_2$ indicate that the kinematics of NGC\,3741 is    
very symmetric. Wong et al. (2004) show that an error in the fitted    
inclination angle causes the $c_3$ term to be proportional to $c_1 ~    
\sin i$, which seems to be the case here. However, as seen in Fig.~\ref{mom0},    
the inclination is quite well constrained from the total HI map. We interpret    
this spurious $c_3 \propto c_1~\sin i$ 
term as the difference between an analysis that assumes    
an infinitely thin disk (the harmonic decomposition of the velocity field)    
and one that takes this effect into account (the data cube modelling). 
Indeed, within the framework of the harmonic analysis this term would 
indicate an incorrect value of the inclination. However, the harmonic
decomposition assumes an infinitely thin disk: if the disk thickness is
actually non-negligible, this assumption
will result in an underestimated inclination angle. Therefore, if one fixes
the inclination at its correct value (accounting for the thickness of the disk),
the output harmonic decomposition procedure will display a spurious 
``$c_3 \propto c_1~\sin i$'' effect.

Note that the plotted errors in Fig. \ref{harmonic} are formal errors coming
from the fitting procedure and they are
likely to be smaller than the real uncertainties in the parameters
(see e.g. Bureau \& Carignan 1999).

\subsection{Model data cubes: high resolution}
\label{section_high}

Starting from the results of the harmonic decomposition of the velocity    
field we built different model data cubes, iteratively improving the input   
parameters of the models until acceptable results were found. 
Some parameters were adjusted at an earlier stage (the position and 
inclination angles), while others (the rotation and radial velocities)
needed some additional small changes.
An automatic and objective method to fit a model data cube to 
an observed one, and to estimate the uncertainties in the fit 
is still a matter of active research (J\'ozsa et al. 2006,
in prep.). However, as a first estimate of the kinematical properties we used the most 
complete objective method available, the harmonic decomposition of the velocity 
field, whose results were improved by matching visually some key features
in the model data cubes to the observed ones.
It is possible to associate a $\chi^2$ to each model but its value cannot
be considered for statistical purposes, it can only be a useful tool
to compare different models. The reasons are that 1) there is no 
objective way to define how big the data cube should be or which portion of the 
cube should be fitted and 2) the clumpiness and inhomogeneity of actual 
distribution of gas cannot be reproduced in the smooth axisymmetric models 
shown here. 

\begin{figure*}   
\begin{center}   
\psfig{figure=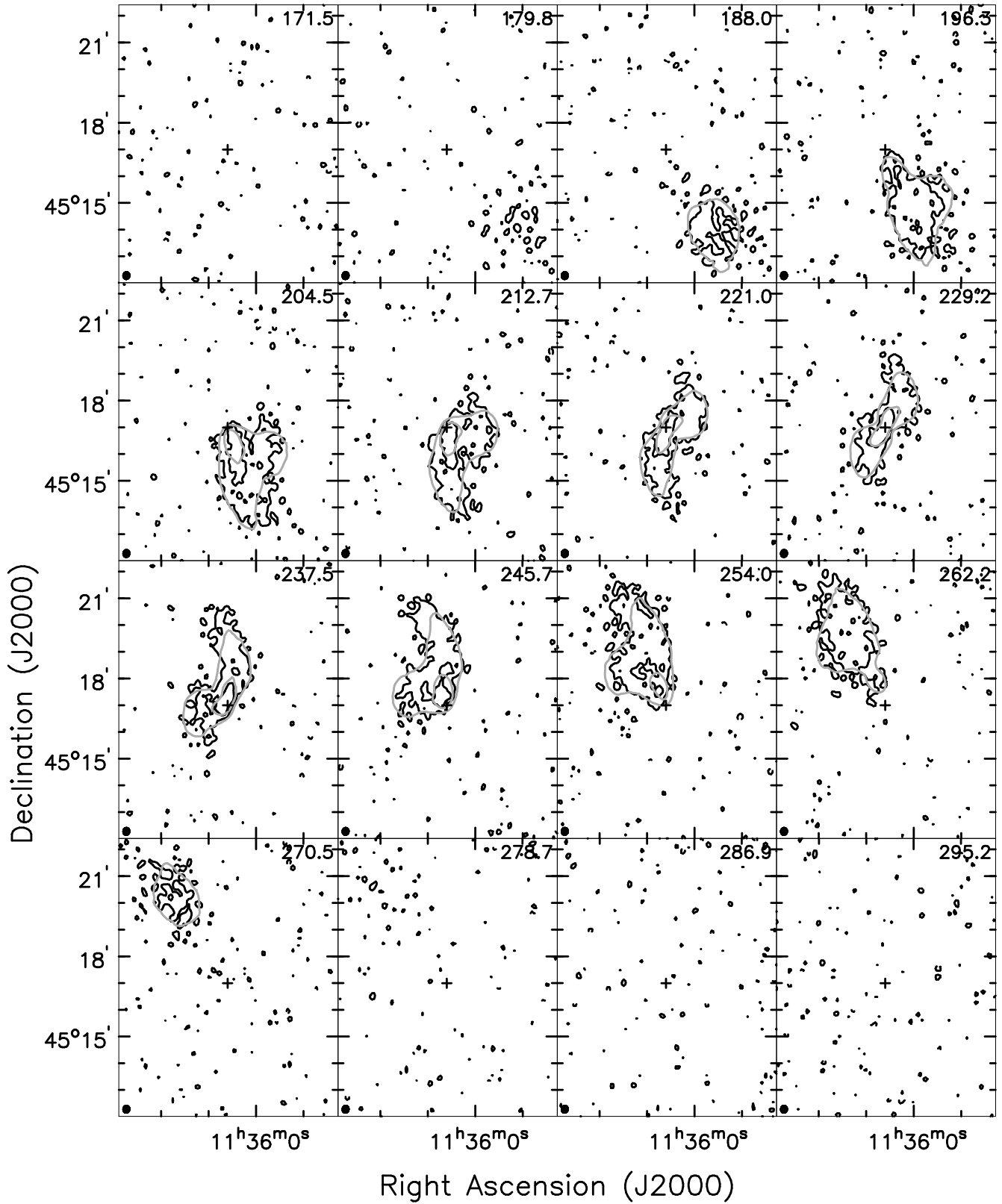,width=17.9cm,angle=0}   
\caption{Comparison between the observed high-resolution data cube (black 
contours) and the modelled one (grey contours). Every second channel map 
is shown. Contours are -2.25, 2.25 (2.5~$\sigma$) and 6.75 mJy/beam.    
The beam is shown in the bottom left corner, and the cross indicates the 
centre of the galaxy. }   
\label{cubes_hires}   
\end{center}   
\end{figure*}

\begin{figure}   
\begin{center}   
\psfig{figure=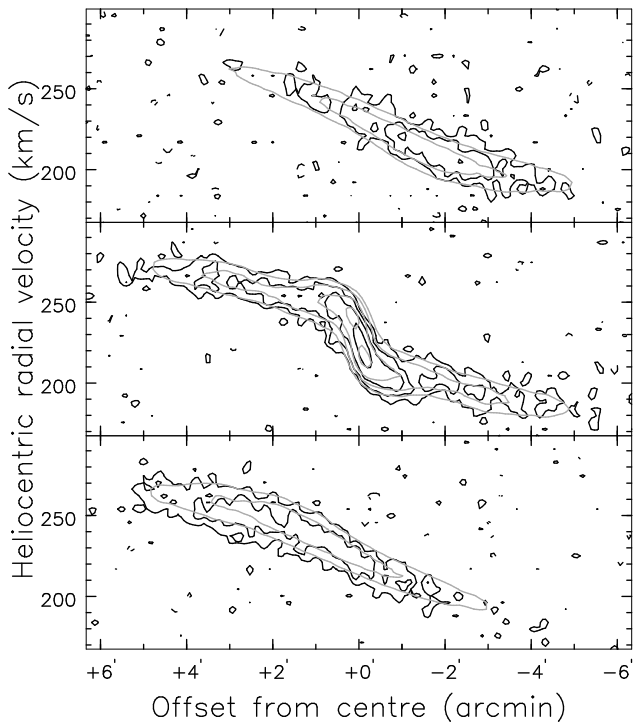,width=8.5cm,angle=0}   
\caption{Comparison between position-velocity diagrams parallel to the    
major axis of the observed high-resolution cube (black contours) and the    
modelled ones (grey contours). The slices are spaced by 70\arcsec. Contours    
are -1.8, 1.8 (2 $\sigma$), 4.5, and 11.2 mJy/beam. }   
\label{pv.3slices}   
\end{center}   
\end{figure}

The rotation curve was derived from the tilted-ring fit of the velocity field   
by only considering those points that are less than 45\degr\ away from the    
major axis, while for the radial velocity as a function of radius the whole    
velocity field was used. It turns out that the value of the rotation and    
radial velocities had to be slightly modified in the inner parts (compared    
to the output of the tilted-ring fit to the velocity field) in order to    
reach a better agreement with the observations. The rotation velocity    
had to be increased by up to 4~\kms\ in the innermost parts, and to be    
decreased by 2~\kms\ in the outer parts of the low-resolution cube. The    
radial velocity was increased by up to 4~\kms\ in the innermost and    
outermost parts. The comparison between our best model high-resolution    
data cube and the observed one is shown in Fig.~\ref{cubes_hires}. With    
our choice of input parameters we are able to reproduce all the major    
features present in the observations.
We reiterate that without the introduction of non-circular motions the
good agreement would not have been reached, due to the impossibility
of reproducing simultaneously the total intensity map (Figs. \ref{mom0_only} and
\ref{mom0_opt_only}) and the central channels (Fig. \ref{cubes_hires_vr0}).
The excellent agreement between    
the two data cubes is also visible in Fig.~\ref{pv.3slices}, in which    
position-velocity diagrams along the major axis are compared. The only    
feature that is not well reproduced is the northern part of the emission    
in channels 237.5~\kms\ to 262.2~\kms\ (receding side, see Fig.~\ref{cubes_hires}).    
However, since the corresponding emission is not present in the approaching    
channels, it means that it is an asymmetric feature of the galaxy, thus it    
is not reproducible with our axisymmetric models.

Bearing in mind the caveats discussed above on the use of the
$\chi^2$ statistics, the fact that our preferred model is a 
better reproduction of the observations
(based on the masked cube, see Section \ref{obsred})
can also be seen in the smaller $\chi^2$ value ($\chi^2=2.45\times 10^4$ for about
2.1$\times 10^4$ degrees of freedom) 
compared to the model data cube built from the first attempt to derive the rotation
curve (Section \ref{velfi_first}, $\chi^2=3.62\times 10^4$) or 
compared to the model with the correct orientation angles but no
non-circular motions (Fig. \ref{cubes_hires_vr0}, $\chi^2=3.97\times 10^4$).

A similar procedure was performed to fit the low-resolution cube (see Section
\ref{model_lores}); the results from the low-resolution cube were used
for radii larger than 180\arcsec, because of its better signal-to-noise
ratio in the outer parts. 

\subsection{Model data cubes: low resolution}   
\label{model_lores}
   
One of the main goals of this paper is to investigate the distribution of   
dark matter at large radii; we smoothed the high-resolution cube to obtain   
a beam of 60\arcsec\ $\times$ 60\arcsec\ in order to highlight the faint    
diffuse emission.

Similarly to the high resolution cube, first a tilted-ring fit on the velocity    
field was performed. Again, the kinematical position angle is different from the    
morphological one, and it needs to be modified in order to reproduce the data    
cube. The same procedure as above was applied to look for the best geometrical   
and physical parameters to describe the observed data cube. The parameters for    
the inner parts were fixed at the values derived from the high-resolution cube;    
based on the signal-to-noise ratio of the high-resolution cube, it was decided    
to only use the high-resolution values up to 180\arcsec. A harmonic decomposition    
of the velocity field was performed, whose results are not shown because they
are similar to  
Fig.~\ref{harmonic}. The $s_1$ term in the outer parts is close to 10~\kms,    
while the non-zero $c_3$ term is again interpreted as the effect of disk thickness;    
the inclination is well constrained from the total HI map (Fig.~\ref{mom0_lores}).

\begin{figure*}   
\begin{center}   
\psfig{figure=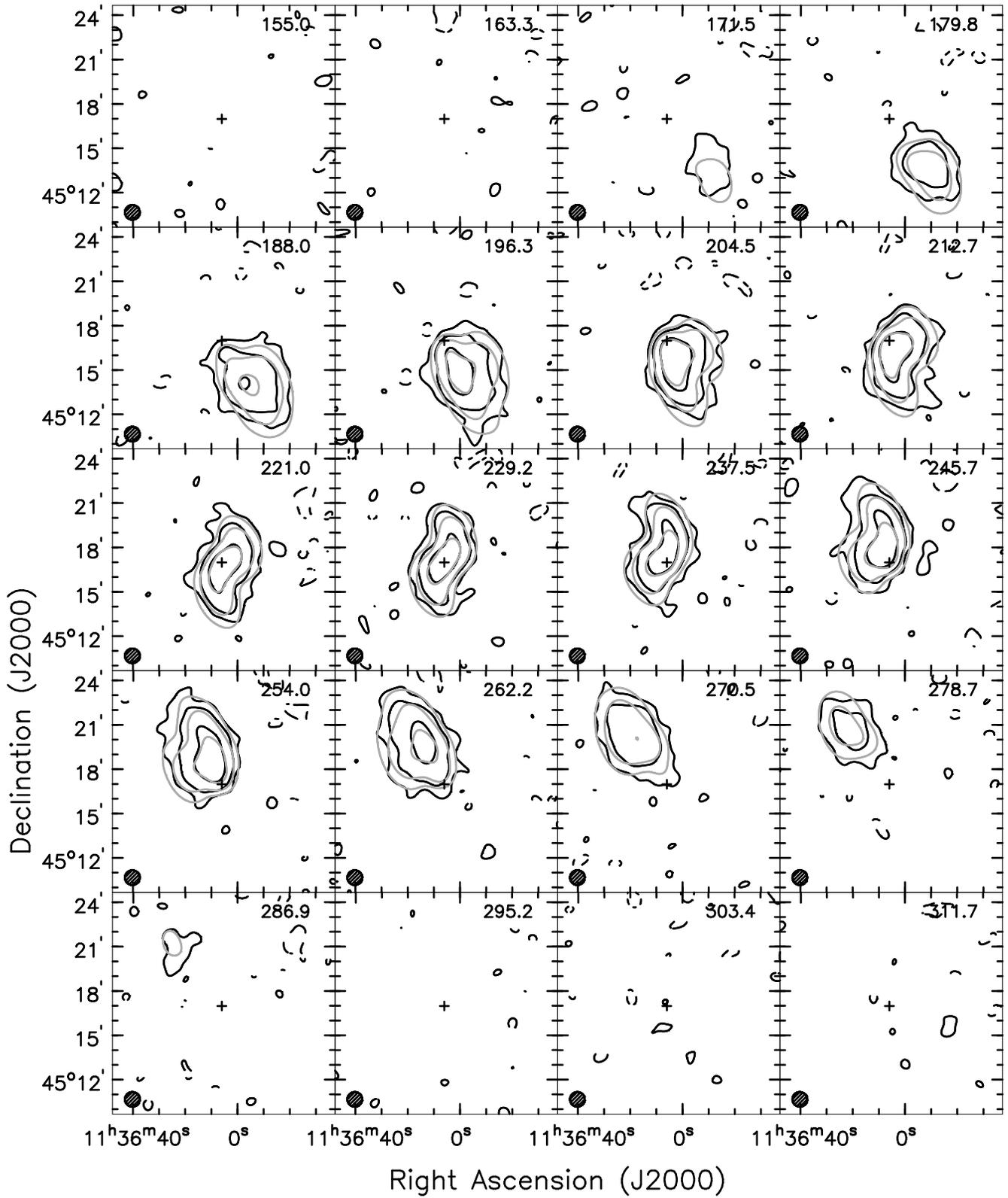,width=17.9cm,angle=0}   
\caption{Comparison between the observed low resolution observed data cube   
(black contours) and the modelled one (grey contours). Every second channel   
map is shown. Contours are -3.25, 3.25 (2.5~$\sigma$), 13, and 52~mJy/beam.    
The beam is shown in the bottom left corner, and the cross indicates the    
centre of the galaxy. }   
\label{cubes_lores}   
\end{center}   
\end{figure*}

\begin{figure}   
\begin{center}   
\psfig{figure=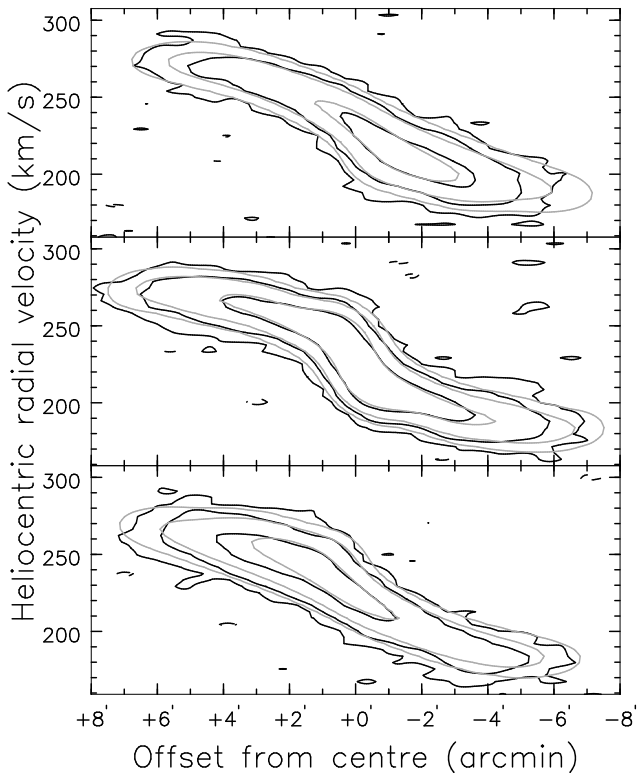,width=8.5cm,angle=0}   
\caption{Comparison between the major-axis position-velocity diagrams of the    
observed low-resolution cube (black contours) and the modelled one (grey    
contours). The slices are spaced by 70\arcsec. Contours are -3.25, 3.25    
(2.5~$\sigma$), 13, and 52 mJy/beam. }   
\label{pv.60.3slices}   
\end{center}   
\end{figure}

The comparison between the observed and modelled low-resolution total HI    
maps is displayed in Fig.~\ref{mom0_lores}. Again, the agreement between    
the maps is excellent. As can be seen in Figs.~\ref{cubes_lores} and    
\ref{pv.60.3slices} our choice of parameters enables us to reproduce the    
observations in detail. Without the introduction of non-circular    
motions such a good agreement would not have been possible. The final choice    
of inclination, position angle, rotation velocity and radial velocity is    
shown in Fig.~\ref{parametri}. The parameters from the low-resolution cube    
were sampled every 30\arcsec\ (half the beam FWHM), so that the low- and    
high-resolution cubes contribute to the final rotation curve with a comparable    
number of points.   
Thanks to the exceptional
extension of the HI disk, the rotation curve extends out
to 6.6 kpc, which correspond to 42 B-band exponential scale lengths
(see Fig. \ref{mom0_opt_only}).

Given the small size of 
the optical disk (R$_{25}$ = 30\farcs7, taken from LEDA, the Lyon-Meudon 
Extragalactic Database) compared to the beam, and given that the orientation 
of the optical disk is not very well defined (e.g. the position angle estimates of the 
optical disk range from $-15.5$\degr, LEDA, to $+23$\degr, Vaduvescu et al. 2005), 
it is hard to define the radius at which the warp commences. According to Briggs 
(1990), warps usually start around R$_{25}$. For the final rotation curve we used 
the values derived from the high-resolution cube out to 180\arcsec, beyond which 
the low-resolution cube was used.

\subsection{Asymmetric drift}

We applied the following correction for asymmetric drift, in order to 
derive the circular velocity $V_{\rm c}$ from the observed rotation velocity 
$V_{\rm rot}$:

\begin{equation}
V_{\rm c}^2=V_{\rm rot}^2-\sigma^2 \left( \frac{\partial~{\rm ln}~ \rho}{\partial~{\rm ln}~ r}
+\frac{\partial~{\rm ln}~ \sigma^2}{\partial~{\rm ln}~ r} \right),
\end{equation}

where $\sigma$ is the one-dimensional velocity dispersion of the gas and
$\rho$ is the gas density, which can be derived from the observed surface density
distribution under the assumption of a constant scale height.
As shown in Section \ref{improved}, the observed velocity dispersion in NGC 3741 is
consistent with a constant value of 6 km s$^{-1}$, which is what we assumed here.
The asymmetric drift correction turns out to be smaller than the error bars except for the 
outermost two data points, where the corrections are respectively 2.7 and 4.8 km s$^{-1}$.
Note that this correction is very uncertain, mainly because we are ignoring the other
gaseous components of the ISM, in particular the molecular gas.

In order to estimate the uncertainties in the 
rotation curve, tilted-ring fits to the velocity fields were made such as those 
illustrated in Figs.~\ref{harmonic}, but considering 
one side of the velocity field at a time. The errors of the final rotation curve 
were taken to be the geometric sum of half the difference between the approaching and 
receding sides and half the asymmetric drift correction.
A minimum error of $2 / \sin i$~\kms\ (half the velocity resolution) was assumed.

\section{Non-circular motions}
\label{section_non}   
   
The distribution and kinematics of the HI in NGC\,3741 could only be modelled   
by resorting to non-circular motions (Sect.~\ref{modelling}). This is    
therefore a caveat in the analysis of the dark matter distribution,    
in that the gas is not in perfect circular motion and the radial component of    
its velocity, especially in the inner parts, is non-negligible. In the inner    
$\sim$30\arcsec\ the non-circular motions that we found are of the same order    
as the rotation velocity. The innermost points of the rotation curve are also    
those that give the strongest constraints on the stellar mass-to-light ratio; therefore,    
the best-fit values inferred in the next section are have larger uncertainties    
than the formal ones, since the rotation velocity $V_{\rm rot}$ is not likely to be a fair tracer of    
the circular velocity $V_{\rm c}$. The non-circular motions as derived from the    
harmonic decomposition are around 10~\kms\ between the centre and 80\arcsec,    
then they drop to about 5~\kms, between 150\arcsec\ and 200\arcsec\ they    
increase again around 10~\kms. As discussed in Sect.~\ref{modelling},    
the higher values were increased by a few \kms\ in order to better reproduce    
the observed data cube, so the range of radial velocities becomes    
5$\ldots$13~\kms. Some dwarf galaxies (such as NGC\,4605, Simon et al. 2005)    
exhibit a similar behaviour in terms of non-circular motions, while in others    
(e.g. DDO\,47, Gentile et al. 2005) their amplitude is much smaller. However,    
the interpretation of these motions is not obvious.   
   
A rough idea can be obtained from the parameters $s_1$ and $s_3$ of the    
harmonic decomposition (Wong et al. 2004). In the inner regions there is a    
hint of an anti-correlation between the two parameters, which can be an indication   
of elliptical streaming. The most likely cause for this kind of motion is an    
inner bar; Figs. \ref{mom0_only} and \ref{mom0} also suggest the presence
of a bar at the centre.
In the outer parts the $s_3$ term is very close to zero, which is    
consistent with a radial flow. In order to disentangle between outflow and inflow    
we need to know the true orientation of the galaxy, i.e. whether it is rotating    
clockwise or counterclockwise. In the map of total HI intensity    
(Fig.~\ref{mom0_lores}) there is a hint of spiral structure. Assuming that    
it is really present and that the spiral arms are trailing, we can infer   
that the galaxy is rotating clockwise, so the radial flow could be interpreted   
as an inflow. 
To try to better understand these radial motions, we separated the galaxy in two parts, 
dividing along the major axis (because the effect of radial motions is stronger along
the minor axis). The radial motions on the north-western side are higher 
(by $\sim$7 \kms) than on the opposite side, which makes the accretion
hypothesis more likely. However, that the magnitude of the motions seems
rather high compared to the theoretical expectations (e.g. Blitz  1996 
expects an inflow velocity of 7 \kms~ in the outer HI disk of the Milky way).
We also note that a full investigation of this phenomenon goes beyond the
scope of this paper. 

\begin{figure*}   
\begin{center}   
\psfig{figure=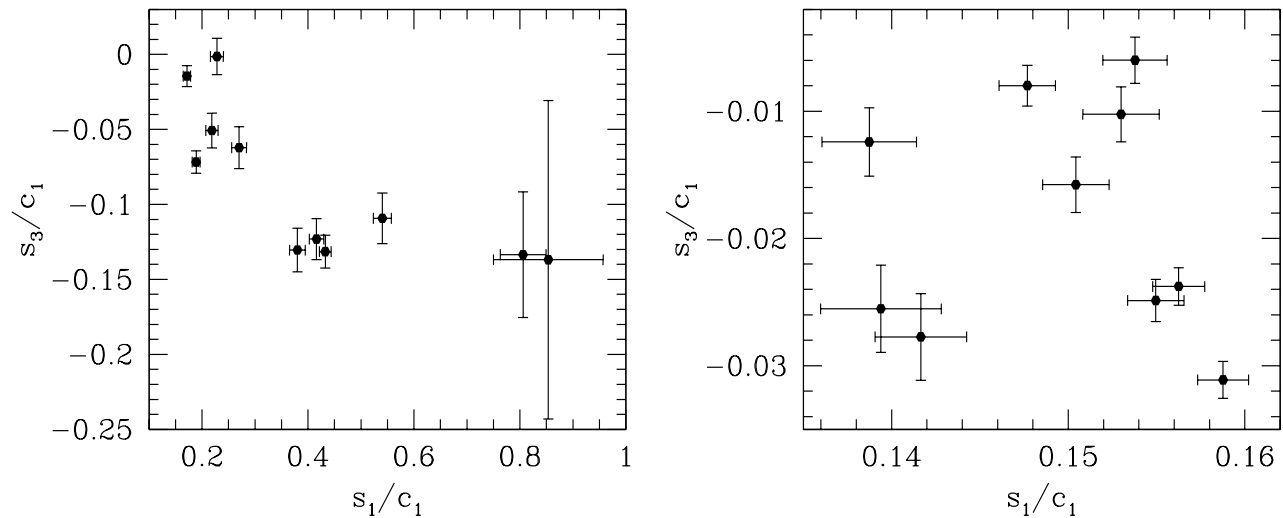,width=17cm,angle=0}   
\caption{
Comparison between the $s_1$ and $s_3$ parameters of the harmonic decompositions
of the velocity fields. Left: for the high-resolution data and radii smaller than
180\arcsec. Right: for the low-resolution data and radii larger than 180\arcsec.
}   
\label{s1s3}   
\end{center}   
\end{figure*}

With this assumption, it might be that the material in the   
outermost regions that is seen having a $\sim 10$~\kms\ inflow is being   
accreted onto the galaxy.
The consistency of the HI observations with an inner bar and outer accretion
can be shown by comparing the $s_1$ and $s_3$ terms: Wong et al. (2004)
showed that they can be use to distinguish between a bar (elliptical streaming),
where the two terms are anticorrelated, and a radial flow, where 
$s_1$ is different from zero and $s_3$ is close to zero.
We plot the two terms (normalised to $c_1$) in Fig. \ref{s1s3}, separately
for the inner regions (radius smaller than 180\arcsec, using the high-resolution
data) and for the outer regions (radius larger than 180\arcsec, using the low-resolution
data). In the inner parts the two terms are anticorrelated, suggesting the presence
of a bar, while in the outer parts they are uncorrelated, which is consistent
with a radial (in)flow.

\section{Mass models}   
   
The data presented in the present paper allow us to make use of the most    
extended rotation curve of a galaxy ever measured, in    
terms of the optical disk scale length.    
The observed rotation curve $V_{\rm obs}$,
corrected for asymmetric drift, was decomposed into    
the stellar disk, gaseous disk and dark halo contributions $V_{\rm disk}$,    
$V_{\rm gas}$ and $V_{\rm halo}$ via   
   
\begin{equation}   
V^2_{\rm obs}(r)=V^2_{\rm disk}(r)+V^2_{\rm gas}(r)+V^2_{\rm halo}(r).    
\end{equation}   
   
The shape of the stellar disk contribution is derived from the NIR photometry    
presented in Vaduvescu et al. (2005). From the K-band photometric profile   
the contribution of the stellar disk was derived using the task ROTMOD in GIPSY
with the assumption of a vertical sech$^{2}$ distribution having a scale height
$z_0=h/6$ (van der Kruit \& Searle 1981). 
Its amplitude is scaled using a constant mass-to-light ($M/L$) ratio. From the    
photometric profile we can exclude any significant contribution from a bulge,    
because NGC\,3741 even shows a brightness deficit at small radii, extrapolating    
the exponential fit to the inner region. The gaseous disk contribution was derived    
from the HI observations presented here, multiplying the neutral hydrogen surface    
density by 1.33 to account for primordial He.

\subsection{Stellar disk}

The stellar $M/L$ ratio being one of the largest uncertainties in 
the rotation curve fits, we describe here the attempts we made to
put some strong a-priori constraints. Unfortunately, with the present data this was not                
possible to achieve. Attempts were made using the measured (B-K) colour and the                  
predictions of stellar population synthesis models of Bell \& de Jong (2001). We                  
find a predicted K-band stellar $M/L$ ratio of 0.3. We also tried to constrain the stellar $M/L$ ratio      
by using the predictions of Bell et al. (2003). They only give values based on the                 
observed B-V and B-R colours. By taking the latter value from LEDA (Lyon-Meudon                  
Extragalactic Database) we obtain a K-band stellar $M/L$ ratio of 0.7.

Another attempt at constraining the stellar $M/L$ ratio was made through
an estimate of the lower limit of the stellar mass by comparing UBV 
photometry (NED) plus JK magnitudes (Vaduvescu et al. 2005) with 
several libraries of SEDs computed with the spectral synthesis
code GRASIL (Silva et al. 1998). The observed SED can be well reproduced 
by several different star formation histories. There are no difficulties in 
producing an amount of about $10^7$ M$_\odot$ or more in stars as
predicted in the cored and in the NFW mass models.  Instead, in order to obtain 
significantly lower masses, i.e. $\leq 5 \times 10^6$ M$_\odot$, 
and a good fit to the observed SED, 
there are some requirements to
be fulfilled: a peculiar SFR, 
a delayed formation for 10 Gyr and most of the stars formed in the last 
0.5 Gyr. In any case, also accepting this fact, by adopting a
standard Salpeter IMF extended from 0.1 to 100 M$_{\odot}$, we
never found that the observed SED implies stellar masses lower 
than $\sim 5 \times 10^6 M_\odot$.
These results, combined with the estimates from the
stellar population synthesis models, 
induced us to consider in the rotation curve fits a minimum K-band $M/L$ ratio of 0.3,
corresponding to a stellar mass of $\sim 7 \times 10^6 M_\odot$ at a distance of 3 Mpc.

\begin{figure}   
\psfig{figure=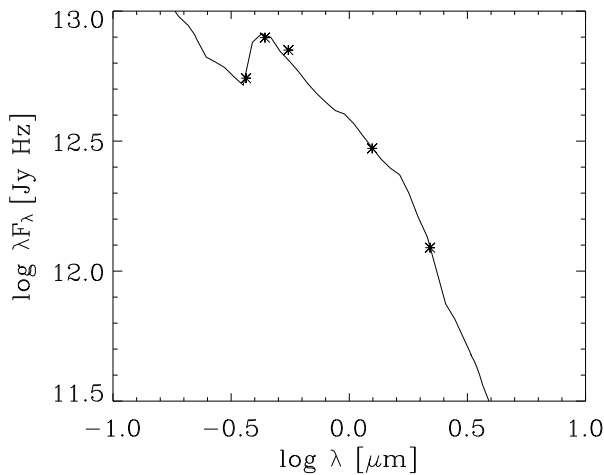,width=8.1cm}   
\caption{
Observed (stars) and fitted (solid line) SEDs, using the GRASIL code
and a stellar mass of $6 \times 10^6 M_\odot$. 
}
\label{grasil}   
\end{figure}

We remind, however, that the stellar
$M/L$ ratio coming out of the fits 
is mostly constrained by the innermost two/three data points of the
rotation curve, which are also the most affected by non-circular motions,
in proportion to the rotation velocity.
In the specific
example shown in Fig. \ref{grasil} the stellar mass is $\sim 6 \times 10^6 
M_\odot$. The spectrophotometric determination of the stellar disk
mass does not help in discriminating 
between cored and NFW halos, and it is not decisive against MOND. 
In any case, the small stellar mass found with these estimates (of the order of
$10^7$M$_{\odot}$) is an advantage for our aims: uncertainties in the stellar
$M/L$ do not have a great influence on the derived dark matter halo profile.

\subsection{Dark halo}   
   
Concerning the dark matter halo, several alternatives were investigated. In    
numerous previous studies dark halos with a central constant-density core    
provided the best fits to the rotation curve. An example of such a cored halo    
is the  Burkert halo (Burkert 1995, Salucci \& Burkert 2001); its density distribution is given    
by    
   
\begin{equation}   
\rho_{\rm Bur}(r)=\frac{\rho_0 r_{\rm core}^3}{(r+r_{\rm core})   
(r^2+r_{\rm core}^2)}   
\end{equation}   
   
\noindent    
were $\rho_0$ (the central density) and $r_{\rm core}$ (the core radius) are    
the two free parameters.    
   
Then, another fit to the observed rotation curve was performed using the    
Navarro, Frenk and White halo (NFW), the result of an analytical fit to the    
density distribution of dark matter halos in CDM cosmological simulations,    
   
\begin{equation}   
\rho_{\rm NFW}(r)=\frac{\rho_{\rm s}}{(r/r_{\rm s})(1+r/r_{\rm s})^2},    
\end{equation}   
   
\noindent    
Here, $\rho_{\rm s}$ and $r_{\rm s}$ are the characteristic density and    
scale of the halo. More recent simulations (Navarro et al. 2004) have    
improved the NFW fit to the $\Lambda$CDM halo density distribution, but    
on scales of radii we are considering here the NFW halo is still a good    
fit to the outcome of the $\Lambda$CDM simulations. The two parameters    
$\rho_s$ and $r_s$ are in principle independent, but they were shown to    
be correlated in simulations (e.g. Wechsler et al. 2002) through the virial    
mass $M_{\rm vir}$, the concentration parameter $c_{vir}=r_{vir}/r_s$ and    
the critical density of the Universe $\rho_{\rm crit}$. The recent
third year WMAP results (Spergel et al. 2006) 
suggest however that the concentration
parameters predicted by Bullock et al. (2001) and Wechsler et al. (2002)
should be lower, in particular due to the lower $\sigma_8$.
As the ``original'' $c_{\rm vir}-M_{\rm vir}$ relation we considered
eq. 18 in Bullock et al. (2001), whose normalisation was lowered by
25\%, compatible with 20\% used in the model of 
Dutton et al. (2006), 28\% that comes out of Bullock et al.'s 
CVIR code and 25\% found by Gnedin et al. (2006). The relations we used were therefore:

\begin{equation}   
c_{\rm vir} \simeq 13.6 \left( \frac{M_{vir}}{10^{11}{\rm M}_{\odot}} \right)^{-0.13},    
~r_{\rm s} \simeq 8.8 \left( \frac{M_{vir}}{10^{11}{\rm M}_{\odot}}   
\right)^{0.46} {\rm kpc}   
\label{cmvir}   
\end{equation}   
   
\begin{equation}   
\rho_{\rm s} \simeq \frac{\Delta}{3}    
\frac{c_{\rm vir}^3}{{\rm ln}(1+c_{\rm vir})-c_{\rm vir}/(1+c_{\rm vir})} \rho_{\rm crit}   
\end{equation}   
   
\noindent    

where $\Delta$ is the virial overdensity at $z=0$ that can be computed
following Bryan \& Norman (1998).
The scatter in the relation that links $c_{\rm vir}$ to $M_{\rm vir}$ has    
been reported to be $\Delta {\rm log} c \sim 0.14$ (Bullock et al. 2001),    
although this value slightly varies from one study to another.   
We also note that these relations assume a running of the tilt of the power
spectrum ($dn_s/dlnk$) of 0, while different values lead to significant
changes in the normalisation of these relations.

\subsection{MOND}   
   
A hypothesis that has been put forward (Modified Newtonian Dynamics, MOND;    
Milgrom 1983) to explain the absence of a Keplerian decline in rotation curves    
is that Newtonian gravity does not hold below a certain acceleration $a_0$,    
without the need of invoking dark matter. The true gravitational acceleration    
$g$ and the Newtonian one $g_{\rm N}$ are linked through:   
   
\begin{equation}   
g=\frac{g_{\rm N}}{\mu(g/a_0)}   
\end{equation}   
   
where $\mu(x)$ is an interpolation function that tends to 1 for $g \gg a_0$   
and tends to $g/a_0$ for $g \ll a_0$. Traditionally $\mu(x)$ was given the   
following form:   
   
\begin{equation}   
\mu(x)=\frac{x}{\sqrt{1+x^2}}   
\label{mond}   
\end{equation}   
   
Recently, Famaey \& Binney (2005) proposed the following alternative form of    
$\mu(x)$, which, contrary to Eqn.~\ref{mond}, Zhao \& Famaey (2006) have shown    
not to be in contrast with the relativistic MOND theory of Bekenstein (2004):   
   
\begin{equation}   
\mu(x)=\frac{x}{1+x} .   
\label{mondnew}   
\end{equation}   
   
In both cases, the stellar $M/L$ ratio and the distance were 
left as free parameters.   

\subsection{Gas scaling}   
   
A noticeable global property of rotation curves of disk galaxies is that the    
ratio of the gaseous disk and dark matter surface density is approximately    
constant with radius (Hoekstra et al. 2001). This implies that one can attempt    
to fit a rotation curve without a dark matter halo by simply scaling up the    
contribution of the gaseous disk. This fact is important in view of the    
hypothesis that a substantial fraction of dark matter within the radii reached    
by rotation curves might reside in a disk, in the form of cold clumps of molecular    
hydrogen (Pfenniger, Combes \& Martinet 1994). Previous studies have shown that    
this hypothesis does not reproduce the details of observed high-quality rotation    
curves of spiral galaxies (Gentile et al. 2004), even though it is a remarkable    
feature of disk galaxies.

\section{Results}   
\label{results}
   
\begin{figure*}   
\begin{center}   
\hspace{0.0cm}
\psfig{figure=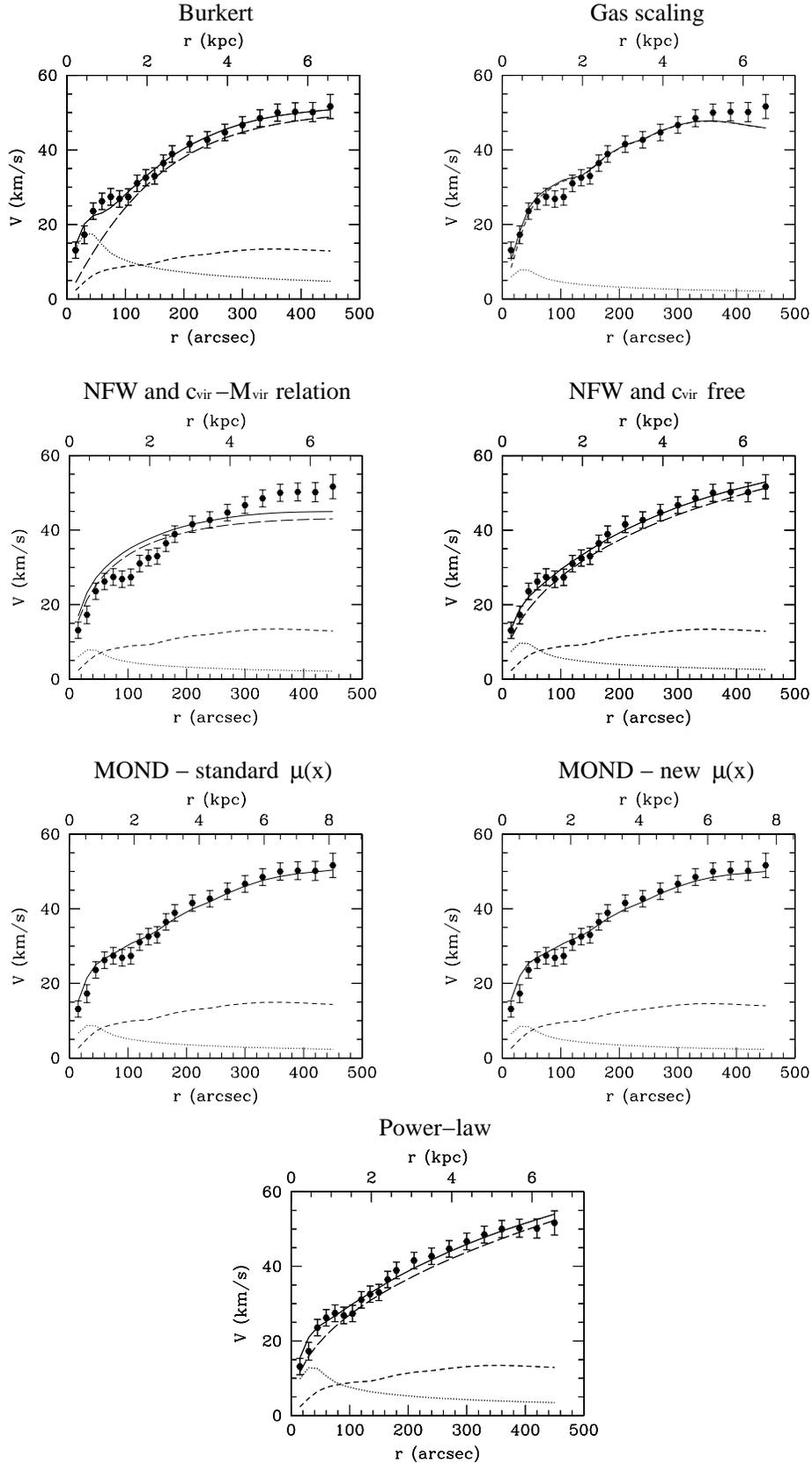,width=12.2cm,angle=0}   
\caption{Rotation curve fits. Dotted/short-dashed lines represent the    
stellar/gaseous disks, the long-dashed line depicts the dark matter halo,    
and the solid line is the best-fit to the rotation curve. The fit labelled   
``NFW and c$_{\rm vir}$-M$_{\rm vir}$ relation'' makes use of Eqn.~\ref{cmvir}    
and the fit called ``MOND - new $\mu(x)$'' uses Eqn.~\ref{mondnew}. 
}   
\label{rcfit}   
\end{center}   
\end{figure*}

The results of the mass decompositions are shown in Fig.~\ref{rcfit} and in    
Table~1. The Burkert halo gives a very good-quality fit, with a core radius    
about five times larger than the optical radius. 
Its best-fit $M/L$ ratio is quite high, rather poorly constrained and 
consistent with the spectrophotometric analysis based on the GRASIL code.
Also, if the $M/L$ in the B-band is considered,     
we find very similar values compared to that found by Carignan \& Purton (1998) for                
DDO\,154, whose properties (e.g. very large extent of the HI disk, maximum rotation                
velocity) are similar to NGC\,3741. 
To further test whether reasonable stellar $M/L$ may weaken the support for a cored halo 
we made a Burkert fit with a fixed $M/L_{\rm K}=1$ (within
the range derived by Bell et al. 2003 for blue galaxies): we find that
the fit is still very good ($\chi^2_{\rm red}=0.62$).

The NFW fit using    
Eqn.~\ref{cmvir} is poor, with the usual effect of overestimating the dark    
mass density in the inner parts and underestimating it in the outer parts.   
Moreover, the best-fit stellar $M/L$ ratio is 0: more
realistic values of $M/L$ would only make the fit worse.

If the relation between $c_{\rm vir}$ and $M_{\rm vir}$    
is relaxed, a good fit can be made. However, this is achieved at the    
expense of rather unrealistic (in the context of $\Lambda$CDM) best-fit    
parameters: the best-fit $c_{\rm vir}$ is 6, which   is about 2.5~$\sigma$ off the    
predicted relation (corrected for the third year WMAP cosmological parameters). 
This means that the rotation curve of NGC\,3741 can    
be fitted with an NFW halo but only if it is a 2.5-$\sigma$ exception to the    
$c_{\rm vir}$-$M_{\rm vir}$ relation. Also, note that in this case the NFW    
best-fit virial mass is too high ($\sim 10^{11}$ M$_{\odot}$) for a galaxy with    
an observed maximum velocity of 45$\ldots$50~\kms, even though the
constraint on $M_{\rm vir}$ is weak.
Moreover, a $M_{\rm stars}/M_{\rm vir} \sim 10^{-4}$ is difficult to explain
even taking the feedback from SN explosions into account.

If the distance is kept as an unconstrained free parameter,
the standard MOND fit is almost as good as the Burkert fit, and so is the MOND fit    
with the new $\mu(x)$ (eq. \ref{mondnew}). In both MOND fits the   
acceleration parameter $a_0$ was kept fixed at $1.2 \times 10^{-8}$ cm s$^{-2}$    
(Begeman, Broeils \& Sanders 1991). 
The best-fit distance in the MOND fits is significantly higher than the value
quoted by Karachentsev et al. (2004), 3.0 $\pm$ 0.3 Mpc,
but it is consistent with the distance
of 3.5 $\pm$ 0.7 Mpc given in Georgiev et al. (1997). The former  
is more likely to be a reliable distance estimate, as it was made using the
tip of the Red-Giant Branch instead of the brightest blue stars,
so the high best-fit distance might potentially constitute a problem
for the standard MOND $\mu(x)$ (see Table \ref{tab-param}). 
By allowing the distance to vary only within the 3.0 $\pm$ 0.3 Mpc range
we found that the standard $\mu(x)$ gives a relatively poor fit ($\chi^2_{\rm red}=1.41$,
which is however still much better than the constrained NFW fit),
while the new $\mu(x)$ gives a better fit ($\chi^2_{\rm red}=0.91$); this fit can
be further improved ($\chi^2_{\rm red}=0.64$) if
$a_0$ is assumed to be 1.35$\times 10^{-8}$ cm s$^{-2}$
(see Famaey et al. 2006). 

Scaling up the contribution    
of the gaseous disk gives a reasonably good fit, with a scaling factor for the    
gas surface density of $\sim 13$, in agreement with the results of Hoekstra et al. (2001).   

A fit of the rotation curve was also made by assuming that the dark matter
density can be described as a power-law (e.g. Simon et al. 2003),
defined as $\rho \propto r^{\alpha}$. 
This is not the case in most models 
of dark matter halos; a power-law fit will therefore provide a sort of average
slope of the dark matter density profile over the range of radii probed by the
rotation curve. Note that the outcome of such a fit depends on the extension of
the rotation curve: a more extended rotation curve will have a steeper average
slope. We fixed the stellar mass-to-light ratio at the average value between the
two extremes found in the above fits: $M/L=0.9$. This quantity turns out not to have 
a great influence on $\alpha$, since the disk
dominates the kinematics only in the innermost few points.
We find $\alpha=-1.11\pm0.11$ and $\chi^2_{\rm red}=0.35$,
which however does {\it not} mean that NGC 3741 has a 
cuspy dark matter halo. Indeed, at their respective core radius, a Burkert halo has 
a $d{\rm log}\rho/d{\rm log}r=-1.5$,
and a pseudo-isothermal halo has $d{\rm log}\rho/d{\rm log}r=-1.0$. This shows
that if the rotation curve extends {\it beyond} the core radius (such as in NGC 3741), 
the average slope, as measured by a power-law fit, will be much steeper than zero.
On the other hand, for an NFW halo it is difficult to reproduce an average slope
close to -1, since its slope is already -2 at $r=r_s$ (which is of the same order
as the last measured radius of our NGC 3741 observations). See e.g. Fig. 3 of Navarro
et al. (2004) for the variation of the density slope with radius 
in $\Lambda$CDM simulations of dwarf galaxies.
The only ways to reconcile an NFW profile with the observations are
to have a very large $r_s$, which means that either the virial mass
has to be very high, or to decrease the concentration,
which is at odds with the $\Lambda$CDM predictions in the
concordance cosmology.

NGC 3741 has therefore evidence for a cored halo. This could be linked to the
presence of a bar. Indeed, barred galaxies seem to have somewhat shallower 
dark matter density profiles (Swaters et al. 2003), despite the uncertainties
in the mass models due to non-circular motions.
However, there is no general agreement yet as to whether realistic bars can 
significantly flatten the cusp (Weinberg \& Katz 2002, Sellwood 2006).

\begin{table}   
\centering   
\begin{scriptsize}   
\centering   
\caption[]{Best-fitting results: parameters and associated 1$\sigma$ uncertainties, reduced    
$\chi^2$. $r_{\rm core}$ is in kpc, $\rho_0$ is in units of    
$10^{-24}$~g~cm$^{-3}$, $M_{\rm vir}$ is in units of M$_{\odot}$,
and the distance is in Mpc. $M_{\rm stars}$ and $M_{\rm gas}$ are the masses of the stellar 
and gaseous disks, respectively, given in M$_{\odot}$. $M_{\rm bar}$=$M_{\rm stars}$+
$M_{\rm gas}$ is the baryonic mass. $M_{\rm vir}$ in the case of the Burkert halo
was computed by extrapolating the density profile until the average density
reached $\Delta~\rho_{\rm crit}$, where $\Delta$ is 92.5.}   
\vspace{0.3cm}   
\label{tab-param}   
\begin{tabular} {l l l }    
\hline     
Burkert                   & $r_{\rm core}$    & 2.97$^{+0.57}_{-0.43}$   \\    
                          & $\rho_0$          & 1.65$^{+0.33}_{-0.33}$   \\    
                          & $M/L$$_{\rm K}$   & 1.51$^{+0.57}_{-0.51}$   \\ 
                          & $M_{\rm stars}$   & 3.47$^{+1.32}_{-1.17}\times 10^{7}$  \\ 
                          & $M_{\rm vir}$     & $\sim$ 6.5 $\times ~10^9$  \\ 
                          & $M_{\rm bar}$/M$_{\rm vir}$ & $\sim$ 3.1 $\times$ $10^{-2}$ \\ \vspace{0.2cm}  
                          &$\chi^2_{\rm red}$ & 0.42                      \\    
Gas scaling               &Scaling factor     & 12.6$^{+0.5}_{-0.5}$      \\    
                          & $M/L$$_{\rm K}$   & 0.30$^{+0.27}_{-0.00}$   \\  
                          & $M_{\rm stars}$   & 0.69$^{+0.62}_{-0.00}\times 10^{7}$  \\   \vspace{0.2cm}  
                          &$\chi^2_{\rm red}$ & 1.00                      \\     
NFW and $c_{\rm vir}$ free&$M_{\rm vir}$      & 1.5$^{+3.9}_{-0.8}\times10^{11}$ \\   
                          &$c_{\rm vir}$      & 6$^{+2}_{-2}$      \\    
                          & $M/L$$_{\rm K}$     & 0.44$^{+0.35}_{-0.14}$   \\
                          & $M_{\rm stars}$   & 1.01$^{+0.80}_{-0.32}\times10^{7}$   \\ 
                          & $M_{\rm bar}$/M$_{\rm vir}$ & $\sim$ 1.2 $\times 10^{-3}$ \\  \vspace{0.2cm}   
                          &$\chi^2_{\rm red}$ & 0.35                      \\     
NFW and $c_{\rm vir}-M_{\rm vir}$   &$M_{\rm vir}$      & 1.4$^{+0.1}_{-0.1}\times10^{10}$ \\     
                          & $M/L$$_{\rm K}$     & 0.30$^{+0.06}_{-0.00}$   \\  \vspace{0.2cm}  
                          &$\chi^2_{\rm red}$ & 4.44                      \\              
MOND - standard $\mu(x)$  &M/L$_{\rm K}$      & 0.30$^{+0.04}_{-0.00}$   \\    
                          &Distance           & 3.71$^{+0.15}_{-0.15}$      \\ 
                          &$M_{\rm gas}/M_{\rm stars}$ & 24$^{+0}_{-2}$ \\ \vspace{0.2cm}  
                          &$\chi^2_{\rm red}$ & 0.52                      \\ 
MOND - new $\mu(x)$       &M/L$_{\rm K}$      & 0.30$^{+0.03}_{-0.00}$   \\  
                          &Distance           & 3.51$^{+0.14}_{-0.14}$      \\ 
                          &$M_{\rm gas}/M_{\rm stars}$ & 24$^{+0}_{-2}$ \\ \vspace{0.2cm}  
                          &$\chi^2_{\rm red}$ & 0.63                      \\ 
\hline   
   
\end{tabular}   
\end{scriptsize}   
\end{table}

\section{Conclusions}   
   
We have presented the analysis of HI observations of the nearby dwarf    
irregular galaxy NGC\,3741, performed at the WSRT. The HI disk of this    
galaxy is very extended, the most extended ever    
observed in terms of optical scale length.
The huge extension of the HI disk enables us to trace the    
rotation curve out to unprecedented distances in terms of the size of the    
optical disk: the last radius at which we can trace the kinematics is 42    
B-band exponential scale lengths, approximately 15 times R$_{25}$.
The HI disk displays a warp, which is very symmetric. 
In order to derive the distribution and kinematics, we built model data    
cubes; with our choice of geometrical and kinematical parameters, the observed    
data cube is reproduced accurately. Some key features in the data cube are    
characteristic of non-circular motions, whose amplitude turns    
out to be of the order of 5$\ldots$13~\kms. Their interpretation is non-trivial,    
but they are consistent with an inner bar and accreting outer material.   
   
Subsequently, the standard rotation curve decomposition was performed.   
The (cored) Burkert dark halo fits the data very well; the fit performed   
by taking the dark matter density distribution predicted by the $\Lambda$CDM   
cosmological simulations fits badly, unless the prediction of a correlation   
between the concentration parameter and the virial mass is relaxed.
In this case, however, the price to pay is a
concentration parameter 2.5 $\sigma$ below the predicted $c-M_{\rm vir}$ relation
and a high virial mass (but poorly constrained) of 10$^{11}$~M$_{\odot}$.

Scaling up the contribution of the gaseous disk    
also gives acceptable fits. MOND, both with the standard interpolation    
function and with the new one proposed by Famaey \& Binney (2005), is also   
consistent with the observed rotation curve. We also notice that the distance 
to NGC 3741 might be a potential problem for the standard MOND interpolation
function, but not for the new one.

\section*{Acknowledgments}   
   
GG wishes to thank Filippo Fraternali for his modified version of GALMOD,   
Ovidiu Vaduvescu for providing the NIR photometric profiles of NGC\,3741
and Gyula J\'ozsa for very useful discussions.   
We thank James Bullock for assistance with his code CVIR, and the 
anonymous referee for his comments.

\end{document}